\documentstyle[epsf,12pt]{article}
\newcommand{\ep}{\epsilon}
\newcommand{\bJ}{\mbox {\boldmath $J$} }
\newcommand{\bJp}{\mbox{\boldmath $J^{\prime}$}}
\newcommand{\btheta}{ \mbox{\boldmath{$\theta$}} }
\newcommand{\bomega}{ \mbox{\boldmath{$\omega$}} }
\newcommand{\bm}{{\bf m}}
\newcommand{\ptheta}{\theta^{\prime}}
\newcommand{\pJ}{{\it J^{\prime}}}
\newcommand{\tJp}{{\it { \tilde J}^{\prime}}  }
\newcommand{\tJ}{{\it\tilde J}}
\newcommand{\ord}{{ \cal O}}
\newcommand{\sa}{{\triangle_{+}}}
\newcommand{\csa}{{\triangle}}
\newcommand{\ssa}{{\frac{\triangle_{+}}{\sigma}}}
\newcommand{\hssa}{{\frac{\triangle_{+}}{\hat{\sigma}}}}
\newcommand{\hsigma}{{\hat \sigma}}
\newcommand{\non}{\nonumber}
\newcommand{\bR}{{\bf R}}
\newcommand{\beq}{\begin{equation}}
\newcommand{\beqa}{\begin{eqnarray}}
\newcommand{\eeq}{\end{equation}}
\newcommand{\eeqa}{\end{eqnarray}}

\begin{document}

\title{Dynamics near Resonance Junctions in Hamiltonian Systems }
\author{Shin-itiro Goto and Kazuhiro Nozaki\\
Department of Physics,Nagoya University,Nagoya 464-8602,Japan}
\maketitle
\begin{abstract}
An approximate Poincare map near equally strong multiple resonances is
reduced by means the method of averaging. Near the resonance junction of
three degrees of freedom, we find that some homoclinic orbits ``whiskers''
in single resonance lines survive and form nearly periodic orbits, each of
which looks like a pair of homoclinic orbits.
\end{abstract}
\pagebreak
\section{Introduction}
\qquad In nearly-integrable Hamiltonian systems of three or more degrees of freedom
, the KAM tori do not partition  phase space into isolated parts and
flows diffuse slowly among different resonant tori
along the Arnold web (resonance web).  This process is known as Arnord
diffusion \cite{arnord}, of which time scale is estimated
 as beyond all order with respect to a small perturbation parameter
\cite{nekho}. The fundamental mechanism for such a
diffusion is believed to be provided by a chain of lower-dimensonal tori
with the transverse intersection of stable and unstable manifolds (
``whiskered tori'')\cite{arnord}\cite{holmes}. Recently, it has been noticed
that the believed picture of the Arnord diffusion is justified only away
from multiple resonances or resonant junctions and the speed of diffusion
is much larger near resonant junctions than that along a single resonance
\cite{laskar}. Near the intersection of a stronger and a weaker resonance
, Haller takes advantage of a near-integrable dynamics and constructs a
faster motions than that of Arnord diffusion along a single resonance
\cite{haller}.  \\
\qquad In this paper, motivated by the above invesigations, we reduce
an approximate Poincare map near equally strong multiple resonances by
means the method of averaging and study the geometry and dynamics near
the resonant junctions.  The following significant property is found for
 a Hamiltonian flow of three degrees of freedom.
The two-dimensional stable and unstable manifolds of a hyperbolic
fixed point of the Poincare map
 intersect transversally at homoclinic orbits, which
 are interpreted as ``survivors'' of ``whiskers'' of two-tori on single
 resonance lines in the Arnold web. Near the homoclinic orbits,
 there are nearly periodic orbits, each of which looks like a pair
 of homoclinic orbits. Some implications of such a pair
 of homoclinic orbits are presented in connection with the Arnold diffusion.\\

\section{Poincare Map near Resonance Junctions}
\qquad Let us consider the following nearly integrable Hamiltonian in the $2N$
dimensional phase space $(\bJ ,\btheta )$, where $\bJ$ and $\btheta$ are
 $N$dimensional action and phase vectors respectively.
\beqa
 H( \bJ ,\btheta )&=& H^{0} ({\bJ}) + \ep H^{1}(\bJ , \btheta ),\non
\eeqa
where $\ep$ is a small parameter and the perturbed Hamiltonian $H^{1}$
is assumed to be analytic in the appropriate domain and given by
\begin{eqnarray}
 H^{1} ( \bJ , \btheta )&=& \sum_{\bm \in {\textbf Z}^{N}} h_{\bm}(\bJ)
  \exp( i \bm \cdot \btheta ) .\non
\end{eqnarray}
The frequencies of the unperturbed Hamiltonian $H^{0}$ are assumed to
satisfy precisely the following ``maximal'' resonance relationships at some
point $\bJ^{0}$ of the action space.
\begin{eqnarray}
\bm^{j} \cdot \bomega^{0} &=& 0 \qquad (1\leq j \leq N-1) ,\non\\
\bm^{N} \cdot \bomega^{0} & \neq& 0 ,\non
\end{eqnarray}
where $\bomega^{0}:= \partial H^{0}/ \partial {\bJ}^{0}$ ~and
$\bm^{j}~(j=1,\cdots ,N)$ are linearly independent integer vectors such that
 the $N\times N$ matrix
\begin{eqnarray}
 M :=({\bm}^{1},{\bm}^{2},..,{\bm}^{N}) , \non 
\end{eqnarray}
 is a unimodular matrix ($\mbox{det} M=1$).
We  introduce the canonical transformation
\beqa
\btheta^{\prime}&=&{M}^{t} \btheta,\non \\
\bJ&=& M \bJp,\non
\eeqa
where $M^t$ denotes a transposed matrix of $M$.
Near the resonance point $\bJ^{0}$ or $\bJp^{0}$ of the action space,
the resonance actions (slow variables) $\pJ_{j} ~(1 \leq j \leq N-1)$
and the non-resonance action $\pJ_{N}$ are scaled as \cite{lichten}
\beqa
\pJ_{j}&=& J^{\prime 0}_{j} + \sqrt{\ep} \tJp + \ord ~(\ep)
\quad (1 \leq j \leq N-1),\non\\
\pJ_{N}&=& J^{\prime 0}_{N} + \ord ~(\ep).\non
\eeqa
In the leading order approximation, the Hamiltonian flow near
the resonance point $\bJp^{0}$ is described by
\beqa
\frac{d \ptheta_{j}}{dt} &=&
\sqrt{\ep} \sum_{k=1}^{N-1} {\tJp}_{k} f_{kj}^{0}+\ord(\ep)
\qquad(1 \leq j \leq N-1), \label{rethe} \\
\frac{d \ptheta_{N}}{dt} &=&\Omega^{0} + \ord(\sqrt{\ep}),\label{nrethe} \\
\frac{d \tJp_{j}}{dt} &=& - \sqrt{\ep}
\sum_{\bm} i m_{j}^{\prime} h_{\bm}(\bJ^{0})~
\exp ~(i\bm \cdot \btheta^{\prime})+\ord~(\ep)
~(1 \leq j \leq N-1), \label{reJ}\\
\frac{d {\tilde J}^{\prime}_{N} }{dt} &=& \ord~(\ep),
\eeqa
where
\beqa
\quad f_{kj}^{0}&:=&\frac{\partial^{2} H^{0}}
{\partial J^{\prime 0}_{k} \partial J^{\prime 0}_{j}},\non\\
\Omega^{0}&:=&\bm^{N} \cdot \bomega^{0}.\non\\
\bm^{\prime}&:=&M^{-1}\bm.\non
\end{eqnarray}
Since $\Omega^{0}\ne 0$ in eq.(\ref{nrethe}), the non-resonance angle
$\ptheta_{N}$ is a fast variable, while  the resonance variables
 $(\pJ_{j},\ptheta_{j}) \quad (1 \leq j \leq N-1)$ are slow variables
as seen in eqs.(\ref{rethe}) and (\ref{reJ}).
Integrating eqs.(\ref{rethe}) and (\ref{reJ}) over a period
$T:=2\pi/\Omega^{0}$ with respect to the fast time so that
the symplectic structure is preserved,
we have the following Poincare map in the leading order approximation:
\beqa
\sa \tJp_{j}(n)&=& -i \sqrt{\ep} T \sum_{\bm} m_{j}^{\prime}(\bm)h_{\bm}
 (J^{\prime 0}) \nonumber\\ &&
\exp(i \sum_{l=1}^{N-1}m_{l}^{\prime}(\bm)
\theta_{l}^{\prime}(n)) \delta _{m_{N}^{\prime},0}\quad (1\leq j\leq N-1),
\label{pmap1}\\
\sa \theta_{j}^{\prime}(n) &=&
\sqrt{\ep}T \sum_{l=1}^{N-1} \tJp_{l}(n+1) f_{lj}^{(0)}
\quad (1\leq j\leq N-1),\label{pmap2}
\eeqa
where $\sa\theta (n) := \theta (n+1)- \theta (n)$ and $n \in {\textbf N}$.
It should be noted that the continuous limit of the approximate
Poincare map (\ref{pmap1}) and (\ref{pmap2}) becomes
 the conventional reduced Hamiltonian flow
 obtained by the method of averaging Hamiltonian
 in the leading order approximation. Although difference between an
 asymptotic expression for the Poincare map and an asymptotic averaged
 Hamiltonian flow is exponentially small in $\ep$ \cite{neish},
 the exponentially small splitting of a homoclinic orbit can not be
 described by the the asymptotic averaged Hamiltonian flow. Therefore,
 we take the Poincare map (\ref{pmap1}) and (\ref{pmap2}) as the
 basic system describing  the dynamics near the maximal resonance
 junctions in this paper.

\section{Symmetric Hamiltonian with $N=3$ }
\subsection{Poincare map}
\qquad As a simple but non-trivial Hamiltonian, we consider $N=3$ and  each degree
of freedom represents the same physical motion, that is, the Hamiltonian
is symmetric with respect to the exchange of a set of canonical variables:
\beqa
H(\bJ,\btheta)&=& H^{0}(\bJ) + \ep H^{1}(\bJ,\btheta),\non\\
H^{0} &=& (J_{1}^{2}+J_{2}^{2}+J_{3}^{2})/4 ,\non\\
H^{1} &=& \sum_{\bm \in {\textbf Z}^{3}}h_{\bm}\exp(i \bm \cdot \btheta),\non
\eeqa
where
\beqa
h_{(m1,m2,m3)}=h_{(m2,m3,m1)}=h_{(m3,m1,m2)}.\label{symet}
\eeqa
Since $H^{1}$ is assumed to be analytic so that $h_{\bm}$ decays
exponentially as $|\bm|$ becomes large, we retain Fourier
components only for
$|m_{1}|+|m_{2}|+|m_{3}|=:|\bm| \leq 4$.
The maximal resonance junction occurs ,for example,
at  $\bJ^{0}=J^{0}(1,1,1)^{t}$, i.e. $\bomega^{0}= \omega^{0}(1,1,1)^{t}$.
Then, linearly independent integer vectors are chosen as
\beqa
\bm^{1}=(1,-1,0)^{t},\bm^{2}=(0,1,-1)^{t},\bm^{3}=(0,0,1)^{t},\non
\eeqa
where $\bm^{1}$ and $\bm^{2}$ generate resonant integer vectors, which
are ,for example, $\bm^{1},\bm^{2},(1,0,-1)^{t},(2,-1,-1)^{t},(-1,2,-1)^{t}
,(-1,-1,2)^{t}$ for $|\bm| \leq 4$.
The Poincare map (\ref{pmap1}) and (\ref{pmap2}) becomes
\begin{eqnarray}
\ssa\tJp_{1}(n)&=&A_{J} F_{J}(\ptheta_{1}(n),\ptheta_{2}(n)),\label{maps1}\\
\ssa\tJp_{2}(n)&=&A_{J} F_{J}(\ptheta_{2}(n),\ptheta_{1}(n)),\label{maps2}\\
\ssa\ptheta_{1}(n)&=&A_{\theta} F_{\theta}(\tJp_{1}(n+1),\tJp_{2}(n+1)),
\label{maps3}\\
\ssa\ptheta_{2}(n)&=&A_{\theta} F_{\theta}(\tJp_{2}(n+1),\tJp_{1}(n+1)),
\label{maps4}
\end{eqnarray}
where
\beqa
F_{J}(\ptheta_{1},\ptheta_{2})&:=&\sin(\ptheta_{1})+\sin(\ptheta_{1}+
\ptheta_{2})+2 {\tilde h}_{2}\{\sin(2\ptheta_{1}+\sin(2\theta_{1}+2\ptheta_{2})\}+\non \\
 &&\qquad {\tilde h}_{3}\{2\sin(2\ptheta+\ptheta_{2})+
\sin(\ptheta_{1}-\ptheta_{2})+\sin(\ptheta_{1}+2\ptheta_{2})\},\non \\
F_{\theta}(\tJp_{1},\tJp_{2})&:=&\tJp_{1}-\tJp_{2}/2 \non\\
\sigma &:=&\sqrt{2\ep T^{2}h_{1}},\non\\
A_{J}&:=&\sqrt{2h_{1}},\qquad A_{\theta}:=1/A_{J},
\non\\
h_{1}&=&h_{(1,-1,0)}=\cdots,~h_{2}=h_{(2,-2,0)}=\cdots,
~h_{3}=h_{(2,-1,-1)}=\cdots,\non\\
{\tilde h}_{2} &:=&h_{2}/h_{1},\qquad {\tilde h}_{3}:=h_{3}/h_{1}.\non
\end{eqnarray}
\subsection{Geometry near Resonance Junction}
\qquad The map (\ref{maps1})-(\ref{maps4}) has a hyperbolic fixed point
$(\tJp_{1},\tJp_{2},\ptheta_{1},\ptheta_{2})=(0,0,0,0)$, which
is the resonance junction. Degenerate eigenvalues of the linearized map at
the fixed point are given by
\beqa
\lambda^{u}&:=&3H\sigma^{2}/4+1+\sqrt{9H^{2}\sigma^{4}+24H\sigma^{2}}/4,\non\\
\lambda^{s}&:=&3H\sigma^{2}/4+1-\sqrt{9H^{2}\sigma^{4}+24H\sigma^{2}}/4,\non
\eeqa
where $ H:=1+4{\tilde h}_{2}+3{\tilde h}_{3}$. The linearized stable ($W^{s}$)
and unstable  ($W^{u}$) manifolds of the fixed point are
\begin{equation}
W^{s,u}=
c_{1}^{s,u}
\left(
\begin{array}{c} E^{s,u} \\ 2E^{s,u} \\ 0 \\1 \end{array}
\right)
+
c_{2}^{s,u}
\left(
\begin{array}{c} 2E^{s,u} \\ E^{s,u} \\ 1 \\0 \end{array}
\right), \label{usmanifold}
\end{equation}
\beqa
E^{s,u}&:=&\frac{2(1-\lambda^{u,s})}{3A_{\theta}\sigma}, \quad
c_{1}^{s,u}~,~c_{2}^{s,u}\in\bR\non
\eeqa

 It is easy to show that the map (\ref{maps1})-(\ref{maps4})
has the following homoclinic orbits $\gamma_{j}~(j=1,\cdots, 6)$,
which connect the hyperbolic fixed point $(0,0,0,0)$ to itself.
\begin{tabbing}
xxxx\=xxxxxxx\=xxxxxxxxxxxx\=xxxxxxxxxxxxxxxxxxxx   \= \kill
{orbit}\>{\quad\it c1:c2}\>{\qquad $\tJp$}\>{\quad $\ptheta$ }\>
{\quad $\tJ$ }\\
$\gamma_{1}$ \> \quad ~1:~1 \>$
\quad\tJp_{1}=\tJp_{2}$\>$\ptheta_{1}=\ptheta_{2}=S_{2}$ \>
$\tJ_{1}=-\tJ_{3},~\tJ_{2}=0 $ \\

$\gamma_{2}$ \> \quad -2:~1\>$
\quad\tJp_{1}=0$\>$2\ptheta_{1}+\ptheta_{2}=0,\ptheta_{1}=S_{2}$\>$
\tJ_{3}=-\tJ_{2},~\tJ_{1}=0$\\

$\gamma_{3}$ \> \quad ~1:-2\>$
\quad\tJp_{2}=0$\>$\ptheta_{1}+2\ptheta_{2}=0,\ptheta_{2}=S_{2}$\>$\tJ_{2}=-
\tJ_{1},~\tJ_{3}=0$\\

$\gamma_{4}$ \> \quad ~1:-1\>$ \quad
\tJp_{2}=-\tJp_{1}$\>$\ptheta_{1}+\ptheta_{2}=0,\ptheta_{1}=S_{1}$
\>$\tJ_{3}=\tJ_{1},~2\tJ_{1}+\tJ_{2}=0$\\

$\gamma_{5}$ \> \quad ~0:~1\>$ \quad
\tJp_{2}=2\tJp_{1}$\>$\ptheta_{2}=0,\ptheta_{1}=S_{1}$\>$\tJ_{2}=\tJ_{3},
~2\tJ_{3}+\tJ_{2}=0$\\

$\gamma_{6}$ \> \quad ~1:~0\>$ \quad
\tJp_{2}=\tJp_{1}$\>$\ptheta_{1}=0,\ptheta_{2}=S_{1}$\>
$\tJ_{1}=\tJ_{2},~2\tJ_{2}+\tJ_{3}=0$
\end{tabbing}

Here, $c1:c2$ indicates the direction of each homoclinic orbit near the
fixed point (see eq.(\ref{usmanifold})),
$J_j=J_j^0+\sqrt{\ep}\tJ_j+\ord~(\ep)$ and $S_1,~S_2$ are
homoclinic solutions  of the following two-dimensinal maps similar to the
standard map.
\beqa
\frac{\triangle^{2}}{\sigma^{2}}S_{2}(n)&=&\frac{1}{2}\{ \sin S_{2}(n)
+(1+2{\tilde h}_{2})\sin 2S_{2}(n)\non \\&&
+3{\tilde h}_{3}\sin 3S_{2}(n)+2{\tilde h}_{2}\sin 4S_{2}(n)\},\label{homoc1}\\
\frac{\triangle^{2}}{\sigma^{2}}S_{1}(n)&=&\frac{3}{2}\{ (1+{\tilde h}_{3})
\sin S_{1}(n)+(2{\tilde h}_{2}+{\tilde h}_{3})\sin2S_{1}(n)\},\label{homoc2}
\eeqa
where $\triangle^{2} S(n):=S(n+1)+S(n-1)-2S(n)$.

Six homoclinic orbits $\gamma_{j}~(j=1,\cdots, 6)$ are classified into
two types according to two homoclinic solutions $S_1,~S_2$, i.e. the type of
$\gamma_{j}~(j=1,2,3)$ and $\gamma_{j}~(j=4,5,6)$ are considered to
be different from each other.

Thus, we obtain a geometrical picture of invariant manifolds near
the resonance junction such that the hyperbolic fixed point at the junction
has two-dimensinal stable and unstable manifolds, of which tangent spaces
are given by eq.(\ref{usmanifold}). The two-dimensinal stable and unstable
 manifolds are shown to intersect transversally at the homoclinic orbits
 $\gamma_{j}~(j=1,\cdots, 6)$ (an infinite number of  homoclinic points)
   , which produce stochastic layers with exponentially small widths
\cite{gelf} \cite{hirata}.
 If we take the continuous limit of (\ref{homoc1}) and (\ref{homoc2}),
the exponentially small effects with respect to $\ep$
are neglected and the above homoclinic orbits are reduced to homoclinic
lines without stochastic layers.
In the next subsection, the homoclinic orbits $\gamma_{j}~(j=1,\cdots, 6)$
are interpreted as ``survivors'' of homoclinic orbits (``whiskers'')
 associated with two-tori in  single resonances.
\subsection{Whiskered Tori in Single Resonances}
\qquad Let us consider a single resonance line connected to the present resonance
junction: for example, set $\bJ^{0}=J^{0}(1,1,r)^{t}$ and
$\bomega^{0}=\Omega^{0} (1,1,r)^{t}$,
where $r$ is an irratinal number near $1$. Then, the resonant integer
vector is $\bm^{1}=(1,-1,0)^{t}$ and
$\bomega^{0}\cdot\bm^{1}=0$, while $\bomega^{0}\cdot\bm^{2}\neq0$ and
$\bomega^{0}\cdot\bm^{3}\neq0$. Since $\ptheta_{1}$ is a only
slowly varing phase, we average eqs.(\ref{maps1})-(\ref{maps4}) over a
fast varing phase $\ptheta_{2}$.  We have $\tJp_{2}=0$ and
\beqa
\sa\tJp_{1}(n)&=&A_J\sigma
   \{\sin\ptheta_{1}(n)+2{\tilde h}_{2}\sin 2\ptheta_{2}(n)\},\non\\
\sa\ptheta_{1}(n)&=&A_{\theta}\sigma\tJp_{1}(n+1),\non
\end{eqnarray}
which is the standard map and has a homoclinic orbit connecting a hyperbolic
fixed point $\ptheta_{1}=0 (~\mbox{mod}(2\pi)), \tJp_{1}=0$.
The two torus characterized by $\tJ_2^0, \tJ_3^0$ is called
a whiskered torus associated with a homoclinic orbit (`` a whisker'')
$(\tJp_{1},\ptheta_{1})$.From the canonical transformation
\beqa
\left(
\begin{array}{c} \tJp_{1} \\ \tJp_{2} \\ \tJp_{3} \end{array}
\right)
&=&\left(
\begin{array}{c} \tJ_{1} \\ \tJ_{1}+\tJ_{2} \\ \tJ_{1}+\tJ_{2}+\tJ_{3}
\end{array}
\right)\non
\eeqa
and $\tJp_{2}=\tJp_{3}=0$, we have
\beqa
\tJ_{1}&=&-\tJ_{2},\quad \tJ_{3}=0,\non
\eeqa
which is the relation of the homoclinic orbit $\gamma_3$ in the
resonant junction. In this sense,
the homoclinic orbit $\gamma_3$ is
interpreted as a survivor of the homoclinic orbit (``whisker'') on the
single resonance line specified by the resonant integer vector
 $\bm^{1}=(1,-1,0)^{t}$.\\
Similar discussions near the other single resonance lines connected to
the resonance junction yield the following correspondence between
the homoclinic orbits $\gamma_{j}~(j=1,\cdots, 6)$ and ``whiskers'' of
two-tori in single resonances.
\begin{tabbing}
xxxxxx\=xxxxxxxxxxxx\=xxxxxxxxxxxxxxxxxxxx   \= \kill
{ orbit}\>{ type }\>{\it \qquad
$\bomega^{0}/\Omega^{0}$}\>{\it \qquad\bm} \\
$\gamma_{1}$ \> $ S_{2}$ \> (1,r,1) \> $(~1,~0,-1)$ \\

$\gamma_{2}$ \> $S_{2}$ \> $(r,1,1)$ \> $(~0,-1,~1)$ \\

$\gamma_{3}$ \> $S_{2}$ \> (1,1,r) \> $(-1,~1,~0)$ \\

$\gamma_{4}$ \> $S_{1}$ \> $(1+r^{\prime},1,1-r^{\prime})$ \>
$(-1,~2,-1) $\\

$\gamma_{5}$ \> $S_{1}$\>$(1,1-r^{\prime},1+r^{\prime})$\>
$(~2,-1,-1)$\\

$\gamma_{6}$ \> $S_{1}$\>$(1-r^{\prime},1+r^{\prime},1)$\>
$(-1,-1,~2)$
\end{tabbing}
All possible ``whiskers'' on single resonance lines connected to
the present resonance junction
survive as the homoclinic orbits $\gamma_{j},~(j=1,\cdots, 6)$.
Therefore, the two different types of homoclinic orbit correspond to
different single resonances, that is  $\gamma_{j}~(j=1,2,3)$ corresponds
to $(1,-1,0)$ etc. and $\gamma_{j}~(j=4,5,6)$ to $(2,-1,-1)$ etc.

\subsection{A Pair of Homoclinic Orbits}
\qquad In order to study dynamics near the homoclinic
orbits in the resonant junction, we set  initial values for
the map (\ref{maps1})-(\ref{maps4}) close to each homoclinic orbit
$\gamma_{j}$ and
obtain  numerical solutions. As shown in Fig.(1), we find a nearly
periodic orbit which looks like a pairing of two types of homoclinic
orbits $\gamma_{1}$ and $\gamma_{4}$.
 Hereafter, such a periodic orbit is called  a pair of
homoclinic orbits for short. Similar numerical calculations as Fig.(1)
indicates that each homoclinic orbit has the definite one as a pair such
that $(\gamma_{1},\gamma_{4}), (\gamma_{2},\gamma_{5})$ and
$(\gamma_{3},\gamma_{6})$. The type of homoclinic orbit in the pair
is different from each other. \\
Implications of such a pair of homoclinic orbits are discussed in the
last section in connection with the Arnold diffusion near resonant
junctions.

\section{Asymmetric Hamiltonian with $N=3$ }
\qquad In stead of the symmetry condition (\ref{symet}), we set
\beqa
h &:=&h_{(1,-1,0)}=h_{(-1,1,0)}=h_{(0,1,-1)}=h_{(0,-1,1)},\non\\
h_{C} &:=& h_{(1,0,-1)}=h_{(-1,0,-1)}\ne h,\non
\eeqa
and $h_{\bm}=0$ for $|\bm|>2$. Then, the Hamiltonian becomes slightly
asymmetric and we have the following Poincare map near the resonance
junction $\bJ^{0}=J^{0}(1,1,1)^{t}$.
\beqa
\hssa\tJp_{1}(n)&=&2h\sin(\ptheta_{1}(n))+2h_{C}
\sin(\ptheta_{1}(n)+\ptheta_{2}(n)),\label{pamap1}\\
\hssa\tJp_{2}(n)&=&2h\sin(\ptheta_{2}(n))+2h_{C}
\sin(\ptheta_{1}(n)+\ptheta_{2}(n)),\label{pamap2}\\
\hssa\ptheta_{1}(n)&=&\tJp_{1}(n+1)-\tJp_{2}(n+1)/2,\label{pamap3}\\
\hssa\ptheta_{2}(n)&=&\tJp_{2}(n+1)-\tJp_{1}(n+1)/2,\label{pamap4}
\eeqa
where $\hsigma:=\sqrt{\ep} T$.
The map (\ref{pamap1})-(\ref{pamap4}) has the same fixed point as that of
(\ref{maps1})-(\ref{maps4}). Eigenvalues of the linearized map at
the fixed point have four distinct values given by
\beqa
\lambda^{u}_{4}&=&1+3\hsigma^{2}h/2+\sqrt{12\hsigma^{2}h+9\hsigma^{4}h^{2}}/2,
\non\\
\lambda^{s}_{4}&=&1+3\hsigma^{2}h/2-\sqrt{12\hsigma^{2}h+9\hsigma^{4}h^{2}}/2,
\non\\
\lambda^{u}_{1}&=&\hsigma^{2}h/2+\hsigma^{2}h_{C}+1+\sqrt{\hsigma^{4}h^{2}+4
\hsigma^{2}h_{C}h+4\hsigma^{2}h+4\hsigma^{4}h_{C}^{2}+8\hsigma^{2}h_{C}}/2,
\non\\
\lambda^{s}_{1}&=&\hsigma^{2}h/2+\hsigma^{2}h_{C}+1-\sqrt{\hsigma^{4}h^{2}+4
\hsigma^{2}h_{C}h+4\hsigma^{2}h+4\hsigma^{4}h_{C}^{2}+8\hsigma^{2}h_{C}}/2.\non
\eeqa
The tangent spaces of the local stable
and unstable  manifolds of the fixed point are two-dimensional vector
space generated by eigenvectors $W^s_j$ and $W^u_j$ associated with
$\lambda^{s}_j$ and $\lambda^{u}_j$ respectively, where

\beqa
W^{u}_{4}&=&
\left(
\begin{array}{c} -E^{u}_{4} \\ E^{u}_{4} \\ 1 \\-1 \end{array}
\right)\non
,\quad
W^{s}_{4}=
\left(
\begin{array}{c} -E^{s}_{4} \\ E^{s}_{4} \\ 1 \\-1 \end{array}
\right) .\non
\eeqa
\beqa
W^{u}_{1}&=&
\left(
\begin{array}{c} E^{u}_{1} \\ E^{u}_{1} \\ 1 \\1 \end{array}
\right)\non
,\quad
W^{s}_{1}=
\left(
\begin{array}{c} E^{s}_{1} \\ E^{s}_{1} \\ 1 \\1 \end{array}
\right) .\non
\eeqa
where
\beqa
E^{s,u}_{4}&:=&\frac{2}{3\hsigma}
( \lambda^{u,s}_{4}-1 ),\quad
E^{s,u}_{1}:=\frac{-2}{\hsigma}( \lambda^{u,s}_{1}-1 ).\non
\eeqa
In this case, we have the following two homoclinic orbits
$\hat\gamma_{4},~\hat\gamma_{1}$.

\beqa
\hat\gamma_{4}&:& \tJp_{1}=-\tJp_{2},\ptheta_{1}=-\ptheta_{2}={\hat S_{1}},
\quad\frac{\csa^{2}}{{\hat \sigma}^{2}}{\hat S_{1}}(n)=
3h\sin{\hat S_{1}}(n).\non\\
\hat\gamma_{1}&:& \tJp_{1}=\tJp_{2},\ptheta_{1}=\ptheta_{2}={\hat S_{2}},
\quad \frac{\csa^{2}}{\hsigma^{2}}{\hat S_{2}}(n)=
h\sin{\hat S_{2}}(n)+h_{C}\sin 2{\hat S_{2}}(n),\non
\eeqa

It should be noted that the directions of $\hat\gamma_{4},~\hat\gamma_{1}$
near the fixed point are identical with eigen-vectors
$W^{s,u}_4,~ W^{s,u}_1$
respectively and $\hat\gamma_{4},~\hat\gamma_{1}$ correspond to
$\gamma_{4},~\gamma_{1}$ in the symmetrical case.\\
If we set initial values at arbitrary points close to the fixed point,
the homoclinic orbit ($\hat\gamma_{4}$) in the direction of the
eigenvector with the largest eigenvalue ($\lambda^{u}_{4}$) is first
observed and the other homoclinic orbit follows
as shown in Fig.(2). Thus, a pairing of homoclinic orbits are also observed
in this case. However, the correspondence between homoclinic orbits in the
resonance junction and whiskers in single resonance lines is restricted.
\begin{figure}[hbp]
\begin{flushleft}
\epsfxsize=14cm
\epsffile{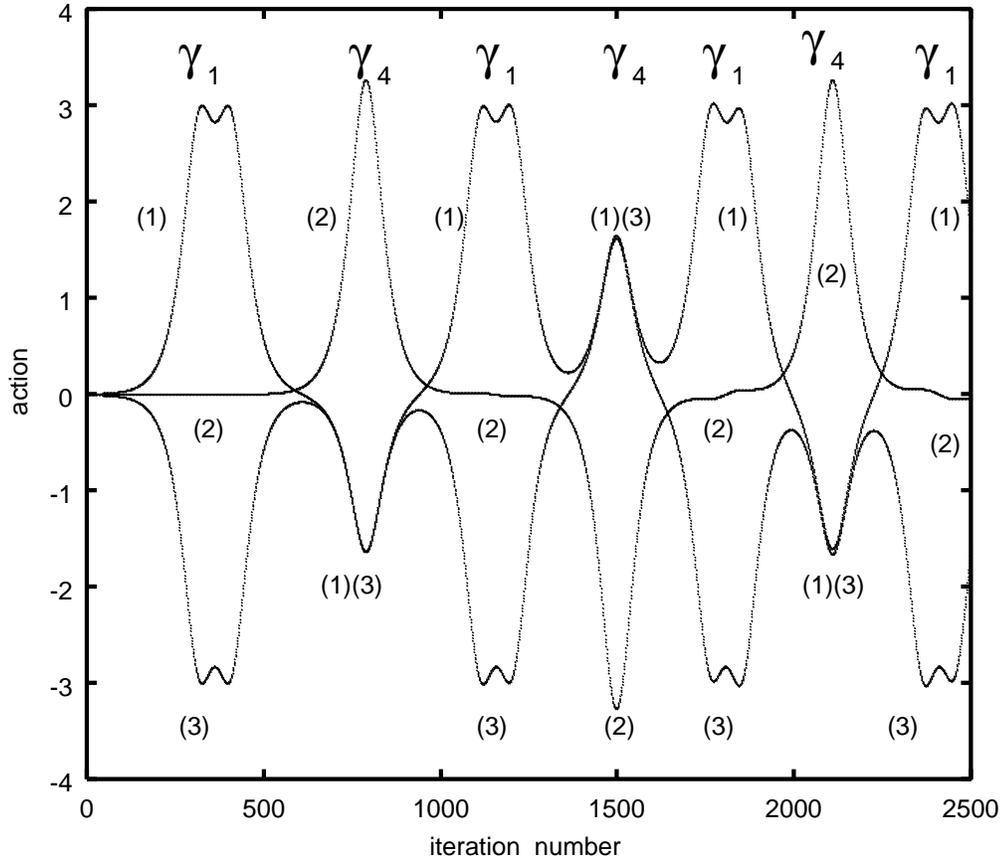}
\end{flushleft}
\caption{A nearly periodc orbit of a pair 
(${\hat \gamma}_{1},{\hat \gamma}_{4}$)for $\sigma=0.02,A_{J}=A_{\theta}=1,
h_{2}=h_{3}=0.001.$ where (1)(2)and (3)denote $\tJ_{1}(n),\tJ_{2}(n)$
and$\tJ_{3}(n)$respectively.}
\end{figure}
\begin{figure}[hbp]
\begin{flushleft}
\epsfxsize=14cm
\epsffile{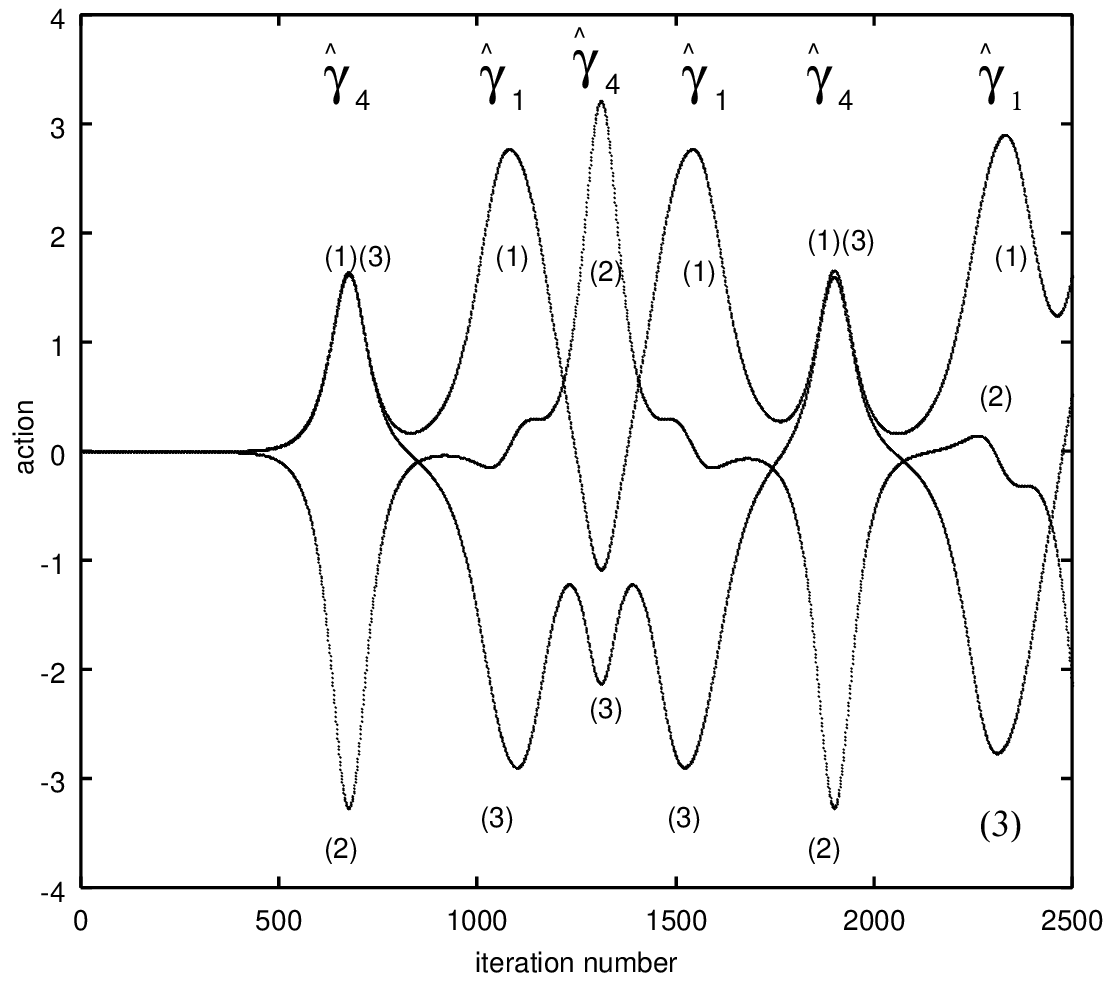}
\end{flushleft}
\caption{
A nearly periodc orbit of a pair 
$({\hat \gamma}_{1}~,~{\hat \gamma}_{4})$ for $\hsigma=0.02,h=0.5,h_{C}=0.1$. 
where (1)(2)and (3) denote $\tJ_{1}(n),\tJ_{2}(n)$ and $\tJ_{3}(n)$
respectively.}
\end{figure}
\pagebreak
\section{Discussion}
\qquad For a symmetrical Hamiltonian system with three degrees of freedom,
 we find the following significant property.
 The maximal resonance junction is a hyperbolic fixed point of the Poincare
 map. The two-dimensional stable and unstable manifolds of the fixed point
 intersect transversally at six homoclinic orbits, which
 are interpreted as ``survivors'' of ``whiskers'' of two-tori on single
 resonance lines in the Arnold web. Near the homoclinic orbits,
 there are three nearly periodic orbits, each of which looks like a pair
 of homoclinic orbits. \\
Let us consider implications of such a pair of homoclinic orbits
in connection with the Arnold diffusion near resonant junctions.
The conventional Arnold diffusion occurs along a single resonance line,
say, $\bJ^{0}=J^{0}{(1,r,1)}^{t}$. The value of irrational $r$ varies very
slowly due to a whisker of $\gamma_{1}$ type. If $r$ reaches $1$,
$\bJ^{0}$ plunges into the resonance junction and  the whisker of
$\gamma_{1}$ type turns into the homoclinic orbit $\gamma_{1}$.
Then, a pairing of homoclinic orbits occurs
and a periodic orbit of homoclinic orbits $(\gamma_{1},\gamma_{4})$
forms in the way shown in Fig.(1). After a long excursion along the nearly
 periodic orbit, the system
will leave the resonant junction and diffuses again along a single
resonance line with a whisker $\gamma_{1}$ or $\gamma_{4}$.
Thus, a pairing of homoclinic orbits at the resonant junction may
give a selection rule for the transision among different single
resonances. 

Although the above implications of a pair of homoclinic orbits
are for the symmetrical Hamiltonian system, similar implications are
provided for a pair of homoclinic orbits $(\hat\gamma_{1},\hat\gamma_{4})$
 in the asymmetrical Hamiltonian system.\\
Furthermore, such a pair of homoclinic orbits will play an important
role in an understanding of dynamics in the Arnold web in general.

\end{document}